\documentstyle[prl,aps,epsf,floats]{revtex}
\def\narrowtext{} \tighten \twocolumn
\input epsf.sty

\begin{document}
\draft

\title{The Fermi surface of Bi$_{2}$Sr$_{2}$CaCu$_{2}$O$_{8}$ }

\author{
        H.M. Fretwell,$^1$\cite{HF}
        A. Kaminski,$^{1,2}$
        J. Mesot,$^2$\cite{JM}
        J. C. Campuzano,$^{1,2}$
        M. R. Norman,$^2$
        M. Randeria,$^3$
        T. Sato,$^4$
	R. Gatt,$^1$ 
	T. Takahashi,$^4$
        and K. Kadowaki$^5$
       }
\address{
         (1) Department of Physics, University of Illinois at Chicago,
             Chicago, IL 60607\\
         (2) Materials Sciences Division, Argonne National Laboratory,
             Argonne, IL 60439\\
         (3) Tata Institute of Fundamental Research, Mumbai 400005, 
             India\\
         (4) Department of Physics, Tohoku University, 980-8578 Sendai, 
             Japan\\
         (5) Institute of Materials Science, University of Tsukuba,
             Ibaraki 305, Japan\\
         }
\address{%
\begin{minipage}[t]{6.0in}
\begin{abstract}
We study the Fermi surface of Bi$_{2}$Sr$_{2}$CaCu$_{2}$O$_{8}$ 
(Bi2212) using angle resolved photoemission (ARPES) 
with a momentum resolution of $\sim 0.01$ of the Brillouin zone. 
We show that, contrary to recent suggestions,
the Fermi surface is a large hole barrel centered at ($\pi,\pi$), 
independent of the incident photon energy. 
\typeout{polish abstract}
\end{abstract}
\pacs{PACS numbers: 71.25.Hc, 74.25.Jb, 74.72.Hs, 79.60.Bm}
\end{minipage}}
\maketitle
\narrowtext

The Fermi surface, the locus in momentum space of gapless 
electronic excitations, is a central concept in the theory of metals.
Despite the fact that the optimally doped high temperature superconductors
display an anomalous normal state with no well-defined 
quasiparticles \cite{KAMINSKI}, many angle resolved photoemission
spectroscopy (ARPES) studies using photon energies in the range of
19-22 eV have consistently revealed a large hole-like Fermi
surface centered at ($\pi,\pi$)\cite{CAMPUZANO,OLSON,AEBI,DING96}
with a volume consistent with the Luttinger count of $(1-x)$
electrons (where $x$ is the hole doping).
This widely accepted picture has recently been challenged by two
studies \cite{DESSAU,SHEN} which suggest a different Fermi
surface when measured at a higher photon energy (32-33 eV).
These recent studies propose that the Fermi surface consists of a large
{\it electron} pocket centered on ($0,0$) with a clear violation
of the Luttinger count.
To reconcile their model with previous data at 22 eV photon energy, these 
authors suggest the presence of ``additional states'' 
near ($\pi,0$), possibly \cite{SHEN} due to stripe formation. 
Setting aside for the moment the important question of what the true 
Fermi surface of Bi2212 is, the implication of a photon energy dependent 
Fermi surface from ARPES data is particularly worrisome, and deserves 
to be addressed.

Here, we present extensive ARPES data taken at various photon
energies and find clear evidence that the Fermi surface measured by 
ARPES is independent of photon energy, and consists of a single hole 
barrel centered at $(\pi,\pi)$.  
Although the data of Refs.~\onlinecite{DESSAU,SHEN}
are consistent with ours, their limited sampling of the Brillouin zone 
and lower momentum resolution lead to a misinterpretation of the
topology of the Fermi surface. This occurs because of the presence 
of ghost images \cite{DING96}
of the Fermi surface due to diffraction of the outgoing 
photoelectrons by a Q vector of $\pm(0.21\pi,0.21\pi)$ associated with
the superlattice modulation in the BiO layers (umklapp bands).
In particular, in following a Fermi contour,
if the data are not dense enough in ${\bf k}$-space, 
or not of sufficiently high momentum resolution, one can 
inadvertently ``jump'' from the main band to one of the umklapp bands, 
concluding incorrectly that the topology of the Fermi surface is 
electron-like. This is particularly relevant at the photon energy of 33eV 
because of a strong suppression of the ARPES matrix elements at {\bf k} 
points in the vicinity of ($\pi,0$), a final state effect, resulting 
in a large umklapp/main band signal ratio near ($0.8\pi,0$) where the 
purported electron Fermi surface crossing occurs.

ARPES probes the occupied part of the electron spectrum,
and for quasi-2D systems its intensity $I({\bf k},\omega)$ 
is proportional to the square of the dipole matrix element, 
the Fermi function $f(\omega)$, and the one-electron spectral
function $A({\bf k},\omega)$ \cite{NK}.  
The measured energy distribution curve (EDC) is obtained by the
convolution of this intensity with experimental resolution.
In another paper, we discuss in great detail the various methodologies 
for determining the Fermi surface from ARPES data\cite{JOELFS}.  
Here, we look at two 
quantities:  (1) the dispersion of spectral peaks obtained from the energy 
distribution curves, and (2) the ARPES intensity integrated over a 
narrow energy range about the Fermi energy\cite{AEBI}.
As we will show,
these methods must be treated with care because of the ${\bf k}$
dependence 
of the matrix elements and the presence of the umklapp bands.

The ARPES experiments were performed at the Synchrotron Radiation
Center, Wisconsin, using a plane grating monochromator beamline with a
resolving power of 10$^{4}$ at 10$^{12}$ photons/s, combined
with a SCIENTA-200 electron analyser used in angle resolved mode. A
typical measurement involved the simultaneous collection and
energy/momentum discrimination of electrons over a
$\sim12^{\circ}$ range (cut) with an angular resolution window of
$\sim(0.5^{\circ},0.26^{\circ})$ ($0.26^{\circ}$ parallel to the cut).
This corresponds to a momentum resolution of (0.038,0.020)$\pi$, 
(0.029,0.015)$\pi$, and (0.022,0.012)$\pi$ at 55, 33, and 22 eV
respectively.
The energy resolution for all data was $\sim$16 meV (FWHM).

\begin{figure}[!t]
\epsfxsize=3.4in
\epsfbox{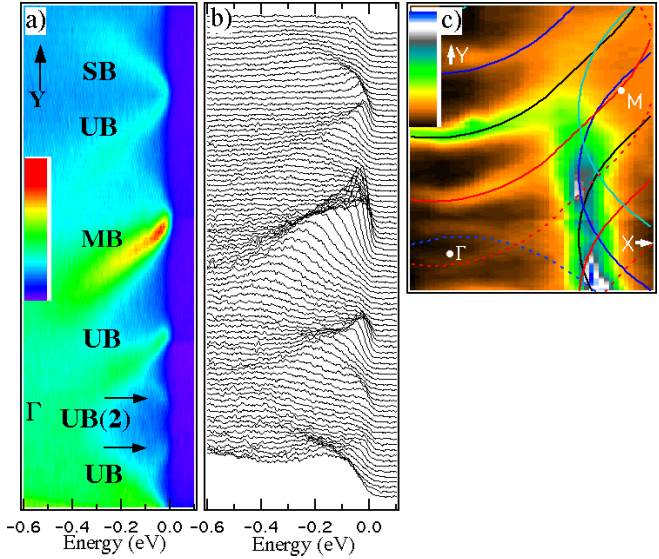}
\vspace{0.5cm}
\caption{
(a) Intensity $I({\bf k},\omega)$ and (b) EDCs along $\Gamma$Y
measured on an optimally doped sample (T$_{c}$=90K) at T=40K with 33 eV
photons polarized along $\Gamma$X.  Main, umklapp, second order umklapp,
and shadow bands are denoted as MB, UB, UB(2), and SB.
(c) Integrated intensity (-100 to +100 meV) covering the X
and Y quadrants of the Brillouin zone. Data were collected on a regular
lattice of ${\bf k}$ points (spacing $1^{\circ}$ along $\Gamma$X and
$0.26^{\circ}$ along $\Gamma$Y).
Overlaid on (c) is the main band (black), $\pm$ umklapps (blue/red),
and $\pm$ 2nd order umklapps (dashed blue/red lines)
Fermi surfaces from a tight binding fit \protect\cite{NORMAN95}.
}
\label{fig1}
\end{figure}

\begin{figure}[!t]
\epsfxsize=3.4in
\epsfbox{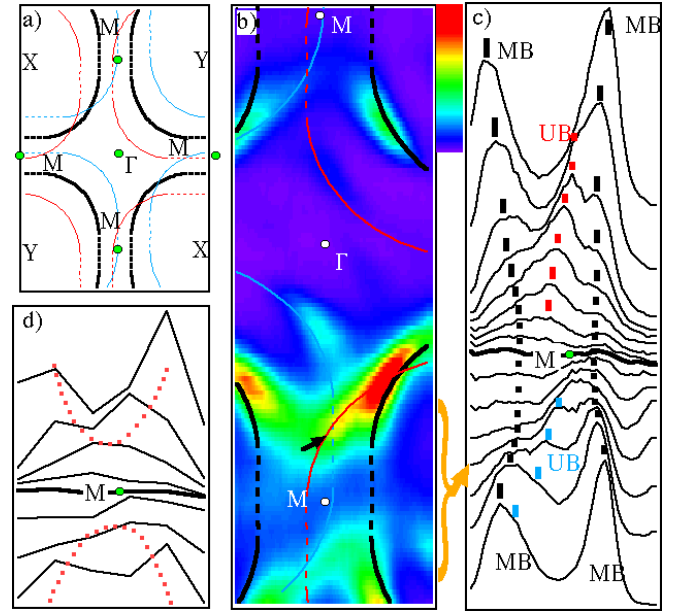}
\vspace{0.5cm}
\caption{
(a) Tight binding Fermi surface from Ref.~\protect\onlinecite{NORMAN95}
and its
umklapp images. 
(b) Integrated intensity (-100 to +100 meV) from four quadrants of the
Brillioun zone. Data measured using 33eV photons on an optimally doped
sample
(T$_{c}$=90K) at T=40K with the same k point spacing as Fig.~1.
Light polarization is along $\Gamma$M. Dashed portions on the Fermi
surface overlay 
from (a) indicate intensity suppresion due to matrix elements. Note 
that because of the umklapps, this leads to a diagonal-like 
suppression of the intensity around M. (c) Slices parallel 
to MY from (b) in an expanded region about M (thick black line is the MY
slice).
(d) Data artificially broadened in ${\bf k}$ resolution by
interpolating (c) onto a new ${\bf k}$ lattice (spacing $2^{\circ}$).
Red dashed lines 
indicate the Fermi surface contours suggested in 
Refs.~\protect\onlinecite{DESSAU,SHEN}.
}
\label{fig2}
\end{figure}

The quality of the optimally doped single crystal samples cannot be
emphasized enough, particularly in regards to the flatness of the
surface after cleave. A change of 1$\mu$m in height over the width of
the sample is readily detectable as a broadening of the spectral
features, and therefore care was exercised in studying very flat
samples with sharp x-ray diffraction rocking curves. Reference spectra
were collected from polycrystalline Au (in electrical
contact with the sample) and used to determine the chemical potential
(zero binding energy).

To begin we look at data, Fig.~1, taken on an optimally doped Bi2212
sample ($T_{c}$=90K), measured at T=40K\cite{SCFS} at 33eV.
The light polarization was parallel to
$\Gamma$X (we use the notation $\Gamma$=($0,0$), X=($\pi,-\pi$),
Y=($\pi,\pi$) and M=($\pi,0$), with $\Gamma$Y parallel to the 
superlattice modulation) and EDCs
were collected on a regular lattice of {\bf k} points ($\delta 
k_{x}=1^{\circ}$, $\delta k_{y}=0.26^{\circ}$). 
We first examine spectra along the
$\Gamma$Y direction. The EDCs are shown in the middle panel of Fig.~1,
and the left panel shows a two dimensional plot of the energy and
momentum distribution of the photoelectrons along the $\Gamma$Y cut.
A strong main band (MB) and additional umklapp bands (UB) can be observed
in this plot.
Around ($0,0$), there is a weaker
pair of higher order umklapps (UB(2) corresponding to a translation of
$\pm(0.42,0.42)\pi$) as observed 
previously \cite{AEBISL}, which confirms the diffraction origin of the 
umklapp bands. Along this cut, we also see the ($\pi,\pi$)
translation of the main band, the so-called shadow 
band (SB) \cite{AEBI}, which is probably \cite{DING96} 
associated with the two formula units per base orthorhombic unit cell.
Fig.~1c shows the integrated intensity within a $\pm$100 meV window
about the chemical potential. 
We note the very rapid suppression of
intensity beyond $\sim0.8\Gamma$M \cite{SAINI}, which does not occur at
22eV. This is what led the authors of Refs. \onlinecite{DESSAU,SHEN} to 
suggest the existence of an electron-like Fermi surface with a 
crossing at this point.
As first discussed in an earlier paper\cite{DING96}, and addressed in 
greater detail here, we will demonstrate that instead, this crossing is
due 
to one of the umklapp bands. This umklapp crossing is more obvious at 33 
eV, since, unlike at 22 eV, the main band intensity is suppressed by matrix
element 
effects\cite{JOELFS}.

To examine this issue more closely, we measured another optimally doped
sample ($T_{c}$=90K, T=40K) with 33eV photons polarized parallel to
$\Gamma$M, 
shown in Fig.~2, with the same high density of
{\bf k}-points as for the $\Gamma$X oriented sample. 
The integrated intensity at
E$_{F}$ is similar to the $\Gamma$X oriented sample in that there is
an `apparent' closed Fermi surface around ($0,0$), indicated by an arrow
in Fig.~2b. However, on closer inspection of the ($\pi,0$) region, we
see what is truly occuring. Fig.~2c shows slices parallel to MY from 
the plot of Fig.~2b. The main band (MB),
indicated by the short black bars, continues to run parallel to
$\Gamma$M,
but its intensity is heavily suppressed near M. In addition, the ($+$)
umklapp band, indicated by red bars, splits away from the main band,
disperses towards M, and dies in intensity. A transposed version of this
occurs
beyond M with the ($-$) umklapp, indicated in blue. 
Similar behaviour is also seen
at the M point at the top of Fig.~2b (slices not shown) and in the
$\Gamma$X
oriented sample of Fig.~1c.

It is easy to see how sparse data at lower resolution can easily lead
one to miss the suppressed main band crossing along MY at 33eV. 
Fig.~2d shows a plot similar to the one in
Fig.~2c, but at a resolution of (0.11,0.11)$\pi$ - the same as that used
in Ref.~\onlinecite{SHEN}- instead of (0.029,0.015)$\pi$ in Fig.~2c.
Clearly, it is no longer possible to distinguish the umklapp from the main 
band (MB),
and one might wrongly suppose that the Fermi surface curves around
to cross the $\Gamma$M line. But we emphasize again that such a
supposition is only a result of sparse data, and not of any inherent
differences in experimental results. It is worth noting that in the
$\Gamma$X
oriented sample (Fig.~1c) we can see a weak signal corresponding to the
main 
band MY Fermi crossing, which becomes stronger at 22eV.
Therefore, if quantities based on integrated intensity are used to
define the Fermi surface, one may falsely infer a crossing along $\Gamma$M
due
to the ($+$) umklapp band, as indicated in Fig.~2d.

The presence of the umklapps can also explain the origin of the
asymmetry in the underlying intensity plot at M in Fig.~2b.
At 33eV the ARPES signal from the main band is strongly
suppressed near M due to the matrix elements,
but since the umklapps are translated by $\pm(0.21,0.21)\pi$, we get
a diagonal-like suppression of the total signal near M. This can be 
appreciated by looking at the dashed segments of the overlay on Fig. 
2b. That is, the umklapp signal at ${\bf k}$ is comparable in intensity 
to the main band signal at ${\bf k \pm Q}$, as expected if the umklapp
is simply a diffraction of the outgoing photoelectrons by the BiO 
superlattice.

\begin{figure*}[!t]
\epsfxsize=6.8in
\epsfbox{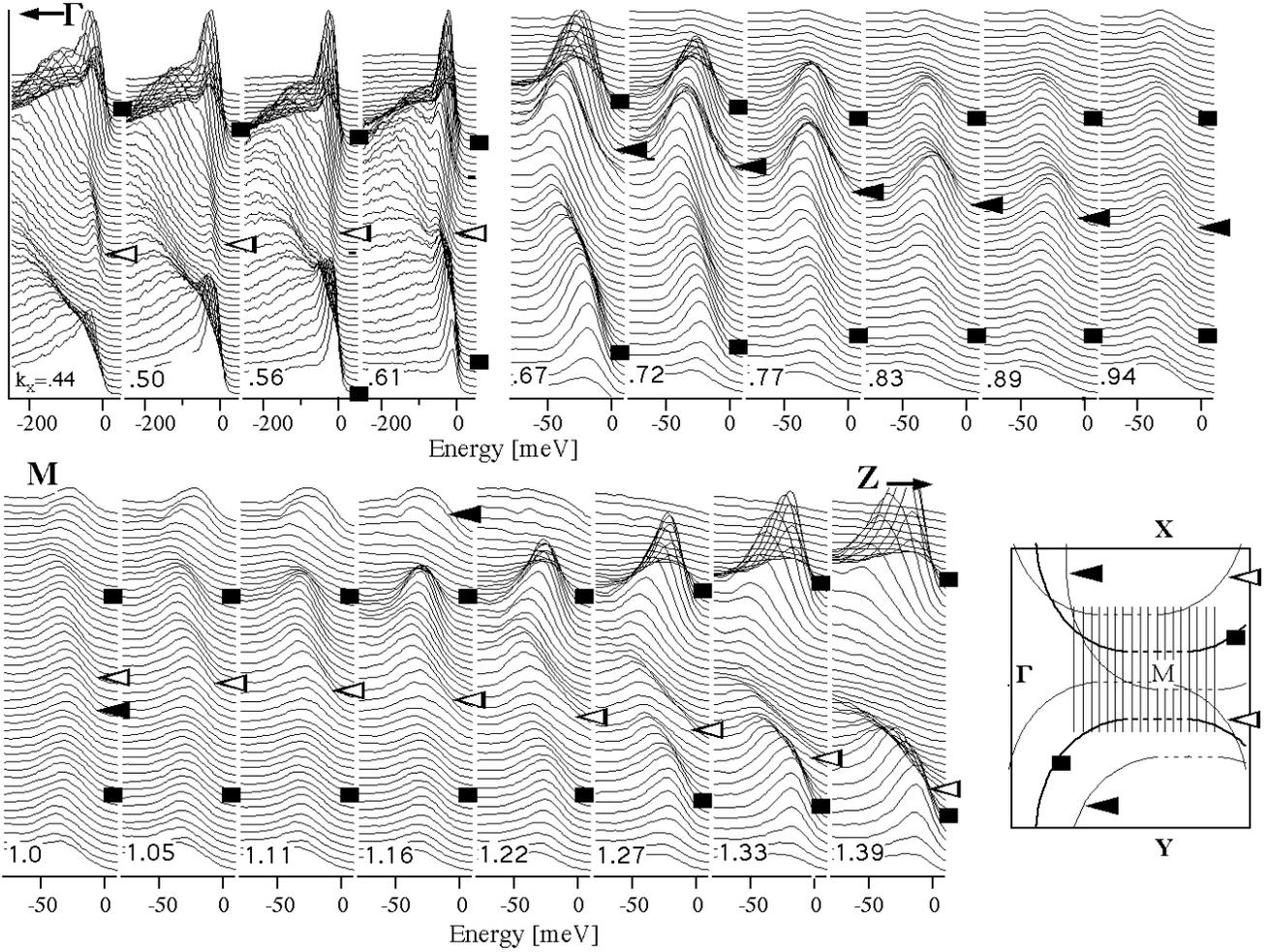}
\vspace{0.5cm}
\caption{
EDCs for cuts parallel to MY (see inset) measured at T=40K for 33 eV
photon
energy.  Note change in binding energy axis scale to emphasize
quasiparticle
dispersion.  Same sample and orientation as Fig.~2. The black squares 
indicate the main band crossings, and the black and white arrows the 
umklapp band crossings.  The zone inset is as in Fig.~2a, with dashed
segments on the Fermi contours indicating matrix element suppression,
and the vertical lines representing the cuts.}
\label{fig3}
\end{figure*}

According to this picture, when moving along $\Gamma$M, there should be a 
crossover from
the ($+$) to ($-$) umklapp, and this is in fact seen in the raw data.
Fig.~3
show extensive EDCs taken in cuts parallel to MY at
{\bf k}-points along $\Gamma$M for 33 eV. In {\it most} of these plots,
the main band (crossing shown by a black square) is the strongest signal
and the $\pm$ umklapps (crossings shown as black and white arrows) are a
weaker
signal superimposed near the $\Gamma$M line.
A constant offset has been used in the figures so that the umklapp
crossings
appear as a ``bunching'' of the spectra. Going from $\Gamma$ to M we see,
in 
the following order, the ($-$) umklapp (white arrow), the ($+$) umklapp 
(black arrow) which disappears at ($\pi,0$), 
and finally the reappearance of the ($-$) umklapp (white arrow). 
We find that the dispersion of all the main and umklapp signals
are consistent with the tight binding fit to the dispersion at
19-22eV \cite{NORMAN95,DING96}. The difference is that at 22eV 
\cite{DING96}, the suppression of the signal at ($\pi,0$) is
weaker, and the MY crossing of the main band is clearer.

\begin{figure}[!t]
\epsfxsize=3.4in
\epsfbox{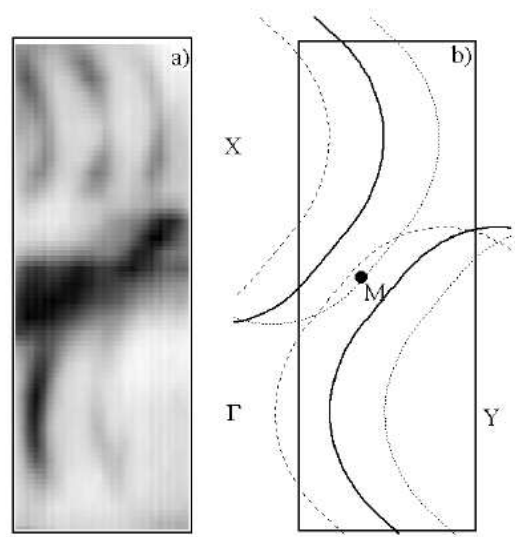}
\vspace{0.5cm}
\caption{
(a) Near $E_F$ intensity (integrated over $\pm$100 meV) at 55 eV for the
$\Gamma$X 
oriented sample of Fig.~1 measured at T=40K.  Note the striking
similarity to 
the large hole Fermi surface (solid curves) with its umklapp images (dashed
curves) shown in (b).
}
\label{fig4}
\end{figure}

In Fig.~4a, we show an intensity map from the sample
of Fig.~1, but at 55 eV photon
energy.  From this intensity plot, one can clearly see the main Fermi
surface
and its two umklapp images, and the correlation of this image with a
single
large hole surface around $(\pi,\pi)$ together with its predicted umklapp
images (Fig.~4b) is striking. 

In conclusion, we find that the Fermi surface of Bi2212 is a single hole 
barrel centered at $(\pi,\pi)$, a result which we find to be independent
of 
photon energy. Rather, we have demonstrated that the unusual intensity
variation 
observed by previous authors at 33 eV is caused by a combination of 
matrix element effects and the presence of umklapp bands caused by the 
diffraction of the photoelectrons from the BiO superlattice. 

This work was supported by the US National Science Foundation,
grants DMR 9624048, and DMR 91-20000 through the NSF Science and
Technology
Center for Superconductivity, the US Dept of Energy Basic Energy Sciences,
under contract W-31-109-ENG-38, the CREST of JST, and the Ministry of
Education, Science and Culture of Japan. 
The Synchrotron Radiation Center is supported by the NSF grant
DMR-9212658.
M.R. was supported in part by the Indian DST through the Swarnajayanti
scheme.

\end{document}